# How the Planned V0 Railway Line Would Increase the Resilience of the Railway Network of Hungary Against Attacks

Bence G. TÓTH,[1] István HORVÁTH[2]

*The spatial distribution of the railway crossings on the river Danube in Hungary is very uneven. There is only one electrified and double-tracked bridge in the country, the Southern Railway Bridge in Budapest. The Újpest bridge in Budapest only provides a connection through line 4 which is not electrified and the Baja bridge is not electrified, too, and both of them are single-tracked. The long-planned V0 railway line that is to be cross the Danube approximately halfway between Budapest and Baja would not only help to redistribute the total network flow which currently passes through almost exclusively the Southern bridge but would also provide redundancy for the existing bridges in the case of their disruption. Four of the five proposed V0 path alternatives are analysed based on these two network properties.*

**Keywords:** *railway, network, disruption, redundancy*

## Introduction

The vulnerability of transportation networks due to highly threatened network elements is a central topic of critical infrastructure analysis. [1] [2] [3] The transportation network of Hungary is heavily centred on Budapest, the country's capital. The main airport of the country with 90% of the traffic is located here, the highways meet in the suburbs and lead to the downtown and the railway lines end at the three main terminals in the city of Budapest. This means that the usage of these networks is not only highly dependent on the urban transport of Budapest but also vulnerable because on the disruption of the common network elements, several transportation lines or even modes are simultaneously disabled.

As the risk of terrorist attacks on the urban railway networks increases, [4] the protection and also the substitution of the highly threatened network elements have to be planned. The dependence of railway lines on each other is the most pronounced at the Southern Railway Bridge in Budapest, which is used by all international freight traffic through

---

[1]  Ph.D., Assistant Professor, National University of Public Service; e-mail: toth.bence@uni-nke.hu; ORCID: 0000-0003-3958-187X

[2]  Ph.D., DSc., Full Professor, National University of Public Service; e-mail: horvath.istvan@uni-nke.hu; ORCID: 0000-0002-1343-1761

Hungary to cross the Danube. This meant 36,445,000 tons of goods equal to 8,589 million ton-kilometres for 2018. [5]

The Southern bridge is the only double-tracked railway crossing on the Danube in Hungary, which is electrified. Since 2018, the Budapest–Esztergom railway line, including the single-tracked Újpest Railway Bridge, is also electrified, but the Esztergom–Almásfüzitő line, which would make it as an alternative of the Southern bridge, is still not.

Furthermore, the transportation tasks of the Hungarian Army also depend highly on the proper condition of the railway network of Hungary, including the sufficient connections between the locations of the troops and also the border of the country. [6] Having only one double-tracked crossing over the Danube, which is the only one which can be used by exclusively electric engines, makes the whole network vulnerable in defence aspects. [7]

## The Defence Requirements of the Railway Network of Hungary

The condition of the Southern Railway Bridge gets poorer every year, therefore, the Government of Hungary decided in 2018 [8] [9] to build a third track which can be used as a replacement while the existing tracks are reconstructed. This third track would also allow increased traffic which would make the whole network depend more on this single network element. [10] [11]

To meet the defence requirements of the nation, the construction of another railway bridge (of another railway line) is necessary. This line could be the long-planned double-tracked electrified V0 railway line. The V0 name is the parallel of the M0 motorway ring around Budapest which connects all the other numbered motorways approaching the city as the V0 line is planned to be a circular line around Budapest exclusively for freight transport that connects the railway lines outside the city. This provides the opportunity for trains with a destination other than Budapest to bypass the few and busy lines (and most of all the critical Southern bridge) inside the capital. [12]

This proposed line was designed by the consortium led by the Association of Hungarian Logistics Service Centres (Magyar Logisztikai Szolgáltató Központok Szövetsége, MLSZKSZ) in 2012 for six alternative routes. However, it was not included in the Integrated Transportation Operative Programme (Integrált Közlekedésfejlesztési Operatív Program, IKOP) [13] and is still not a Government priority for transportation development plans. [14] This is mainly because the Budapest Intermodal Logistics Centre (Budapesti Intermodális Logisztikai Központ, BILK), the main container transfer station was built inside Budapest and therefore entering the city is necessary for all freight trains to exchange their load. Furthermore, a network of intermodal container terminals was also decided to be built. [15]

However, the V0 plan is reintroduced from time to time by railway development specialists to the Government, last time in 2017 by the Hungrail, Hungarian Railway Association [16] as this would promote the establishment of new logistics centres via the advantages of the simultaneous presence of motorways and railways. Furthermore, the important role the V0 railway line could play in the defence capabilities of the country is also an important aspect when considering the necessity of its construction.

Therefore, the aim of the present study is to analyse the proposed paths of the V0 railway line in two manners: which alternative would redistribute the paths of the trains more evenly in the network and which alternative would provide the most redundancy for the existing Danube bridges, i.e. which one is the most sufficient for the defence preparation of the country.

## *The V0 Alternatives*

Six alternative routes for the planned V0 was proposed by the MLSZKSZ (see Figure 1). [17] Route #6 was, however, only the maintenance of the existing lines while Route #5 is not only very similar to route #4 but also too close to Budapest though one of the goals of the project is to relieve the suburbs of Budapest from the heavy rail freight traffic. Therefore, only the analysis of routes #1–#4 was carried out.

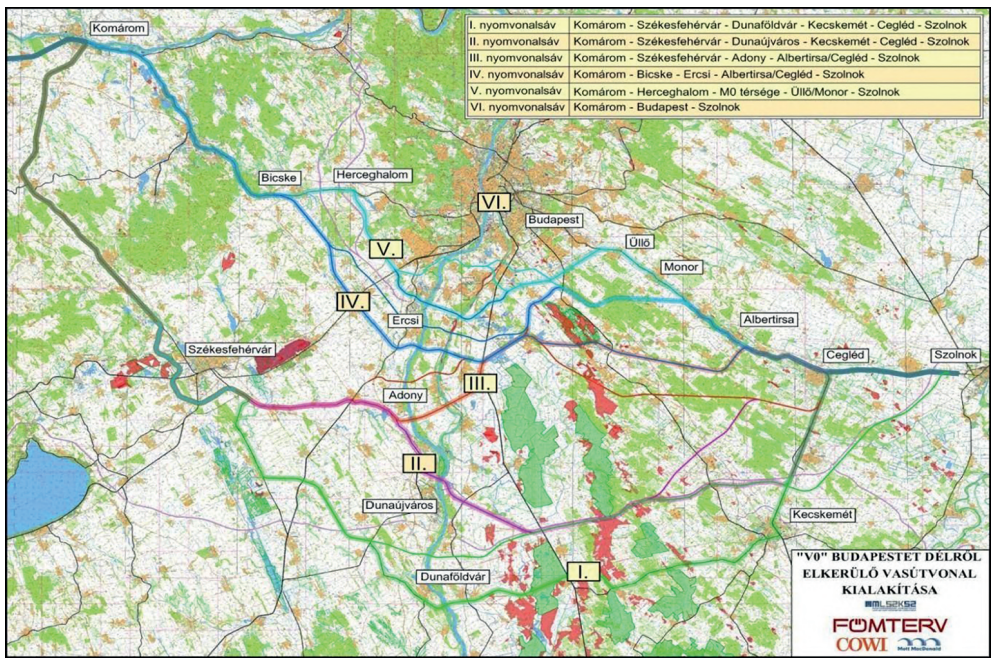

Figure 1. *The six alternative routes for the V0 railway line proposed by the MLSZKSZ.* [17]

## **The Graph Model of the Railway Network of Hungary**

The railway network of Hungary is a 7,441 km long standard gauge system. This means 8.00 km/100 km$^2$ density, [18] which is one of the densest in the world. [19] A weighted graph is used to model the network. [20] The nodes of the graph are the stations and the termini, i.e. where reversing of the trains is possible. Thus, stops with no switches were not included in the model. The edges of the graph are the line sections between the stations.

Two weights can be assigned to each edge, either the length of the travel time of the corresponding line section. The travel time of a line section is calculated as the ratio of the length and line speed. When the line speed is different for different axle loads or trains with locomotive and EMUs, the speed for the highest axle load was taken and a train with a locomotive was assumed, i.e. the smallest line speed value was used. The length and speed data are available online at the page of the Hungarian Rail Capacity Allocation Office (Vasúti Pályakapacitás-elosztó Kft.). [21] The data for the sidings of the Hungarian Army is from a Government Decree. [22] Other parameters of the infrastructure, such as the number of tracks, electrification, maximum allowed axle loads, temporary and permanent speed restrictions are not included in the model.

The stations with exactly two neighbouring stations, the so-called joint nodes, are also transformed out: each joint node and its two connecting edges were substituted by a single edge with a weight of the sum of the two edges replaced. [23] [24] The only exceptions were the stations at which there are no sidings of the Hungarian Army but the nearest towns are the dislocations of a troop.

In total, the railway network was reduced to 292 stations (including termini) and 364 line sections. Within Budapest, four extra nodes were added to the graph to ease further addition of edges related to planned development studies. These four nodes were not regarded as stations, i.e. they were neither origins nor destinations of any path. [25]

The model was developed to find the shortest paths either in distance or time between two arbitrary stations for freight trains with locomotives. This led to the need for adding extra time at the station(s) where locomotive reversal is necessary, for which 15 minutes was assigned. No extra trip length or travel time was assigned to passing a station and no extra distance was assigned to reversing. For this purpose, the graph describing the network had to be modified for the algorithm calculating the shortest path to add the extra time of reversing when needed.

To achieve this, each station was represented by four nodes instead of one: an arrival and a departure node at each side of the stations and four directed edges between them, with the two edges connecting the arrival and the departure node on the same side of the station weighted with 15 minutes. The realisation can be seen in Figure 2.

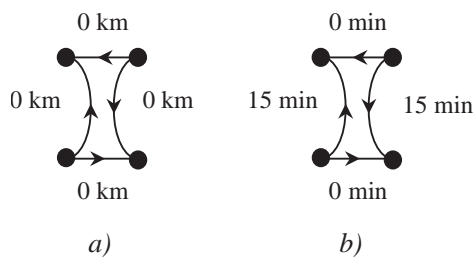

Figure 2. *The weighting of the edges between the four nodes representing a station in the case of calculations of a) trip lengths and b) travel times.* [Made by the authors.]

However, this is not enough, as a path arriving at a station at its arrival node can continue to the next station without passing through the 15-minute edge. To prevent this, the edges between the stations have to be doubled (with the same weight for both) and directed, each pointing from a departure node to an arrival node (see Figure 3/a). This arrangement ensures the addition of the extra time of the reversing.

Furthermore, there are seventeen stations in the network which can be bypassed using a wye (or, in some cases, multiple wyes), which were also represented in the graph with four nodes, similarly to the stations. Since reversing on wyes without entering the corresponding station is not possible, the edges connecting the arrival and departure nodes on the same sides of these quasi-stations were not included (see Figure 3/b).

Therefore, the 289 + 4 stations (plus the wyes) and the 361 line sections were represented by a total of 1,136 nodes and 1,810 edges in the graph.

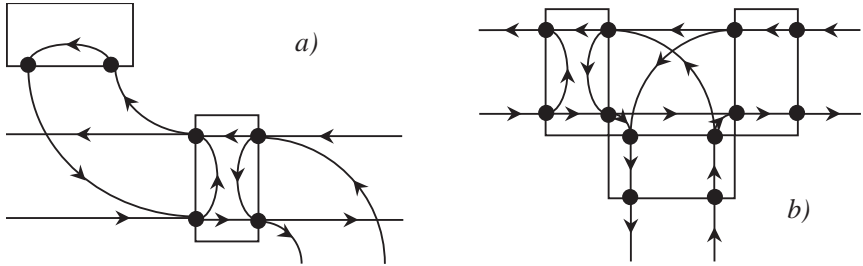

Figure 3. *The representation of termini (a) and wyes (b) in the graph.*
[Made by authors.]

## Methods and Measures

### *Calculating the Shortest Path*

For the calculations and the visualisation, the *igraph* package developed by Gábor Csárdi and Tamás Nepusz [26] in the *R* programming Language and Environment [27] was used. The *igraph* package handles graphs as a set of edges, a so-called edge list: an edge is defined by the two nodes it connects (and optionally, a weight).

The shortest path between two stations in kilometres or minutes was calculated using the *distances()* function of the *igraph* package. For weighted directed graphs with no negative weights (as in our case), the function uses Dijkstra's algorithm [28] as a default.

Calculating the shortest path between all ⟨a,b⟩ pairs of stations with both the time and length weights, 41,616 trip length values and the same number of travel time values are obtained. The *shortest_paths()* function of the *igraph* package makes it easy to determine whether a path passes through a specific line section or station as the function gives four values as a result, two of which are lists, *$epath* and *$vpath*, which list the *ids* of the edges and nodes on the shortest path, respectively. The *id* of an edge is its position in the edge list, while the *id* of a node is the number the node is referred to in the edge list. From these lists, the one with the line section *ids* is needed for the calculations presented.

## *Flow in the Network without the V0*

Calculating the exact shortest paths between all ⟨*a,b*⟩ pairs of stations both for distance and time weights, an artificial flow on the network is obtained (Figure 4).

By summing the trip length and travel time values for all shortest paths, the so-called total network trip length ($c_\ell$) and total network travel time ($c_t$) are obtained. For the railway network of Hungary, in its undisrupted state, these values are $c_\ell$ = 10,047,606 km and $c_t$ = 6,834,569 min = 12.99 year.

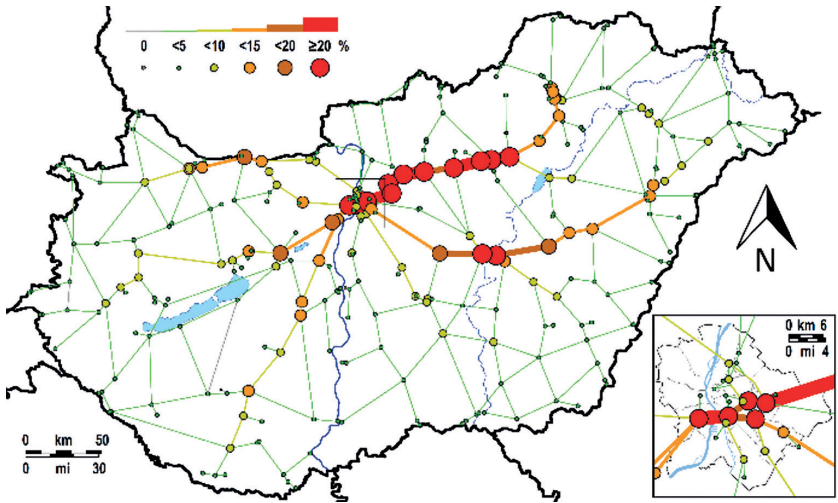

(a)

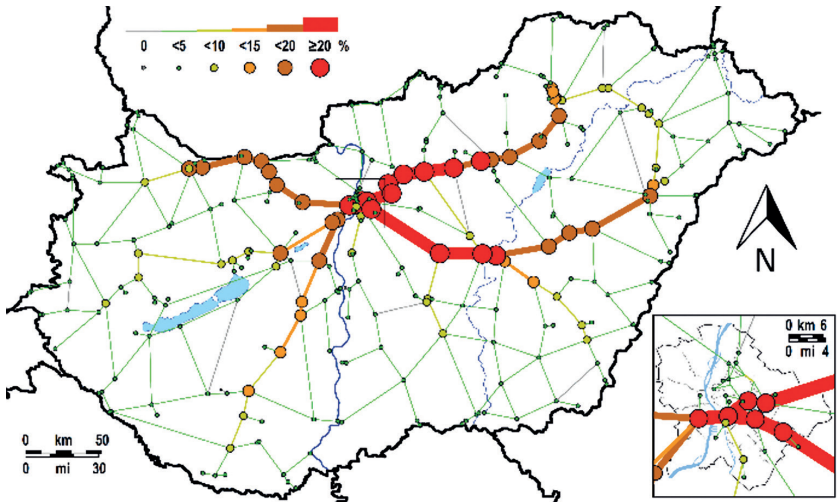

(b)

Figure 4. *The traffic flow values used in the analysis for paths with shortest trip lengths (a) and shortest travel times (b).* [29]

## *Quantifying the Effect of Disruptions*

The term "disruption" will be used in the meaning that the given line section is not available for traffic at all, i.e. no shortest path can pass through it. This was realised by deleting the two corresponding edges of the appropriately weighted graph.

The Network Robustness Index

A commonly used measure to describe the overall resilience of a network against disruptions is the so-called Network Robustness Index (NRI) introduced by Scott et al. [30] The NRI can be calculated for all edges of the graph based on which the importance of the individual line sections can be determined.

To calculate the NRI for line section $u$, the shortest paths between all pairs of stations in the undisrupted graph has to be determined. Then, the total network trip length or the total network travel time is calculated. As the calculation of the NRI does not depend on the weighting, instead of $c_\ell$ and $c_t$, simply $c$ is used to denote this measure.

Then, the edges representing line section $u$ are deleted from the graph. Again, the shortest paths between all pairs of stations are determined and their lengths or durations are summed. This value is denoted by $c^u$. The NRI is calculated as the difference between these two values and is denoted by $q^u$:

$$q^u = c^u - c. \tag{1}$$

The difference is made in this order for $q^u$ to be non-negative since for most kinds of weights (as in the case of length and duration) the deletion of a line section makes $c^u$ larger than $c$ (or at least does not make it smaller). This can be done for all line sections or multiple line sections. If line sections $u$ and $v$ are simultaneously deleted, the NRI is calculated as

$$q^{uv} = c^{uv} - c \tag{2}$$

where $c^{uv}$ is the total network trip length or the total network travel time without the two pairs of edges corresponding to line sections $u$ and $v$. The value of $q^u_{ab}$ (the difference in the shortest path between stations $a$ and $b$ in the disrupted and the undisrupted network) shows if the shortest path in the undisrupted network passes through line section $u$. If $q^u_{ab} = 0$, then line section $u$ is not part of the shortest path between stations $a$ and $b$ neither in the undisrupted nor in the disrupted network. If $q^u_{ab} = 0$, then by deleting line section $u$, the length or duration of the shortest path between station $a$ and $b$ increases compared to the shortest path in the undisrupted network. This means that line section $u$ was part of the shortest path in the undisrupted network but there is still a non-infinite route between stations $a$ and $b$ in the disrupted network.

The Redundancy Index

The Network Robustness Index measures the increase in the total network trip length or the total network travel time in the case of the deletion of a line section. But on the disruption of line section *v*, the exact route of the shortest path between stations *a* and *b* changes compared to the shortest path in the undisrupted network.

Let us assume that the shortest path between stations *a* and *b* in the undisrupted network did not pass through line section *u* but in the disrupted network it does. How much would be the additional increment in the shortest path if *u* would be deleted, too? In other words, how larger would $q_{ab}^u$ be than $q_{ab}^u$, i.e. we want to know how much total increase is caused by deleting not only line section *v* but also line section *u* for those paths that did not pass through line section *u* in the undisrupted network but did pass through in the network without line section *v*. This increase is the redundancy provided by line section *u* to line section *v*. Paths that do not pass through line section *u* in graph $G^0$ or in graph $G^v$ or do pass through it in both graphs are not relevant, since they are not sensitive for the disruption of line section *u*.

Therefore, only those shortest paths are taken into account for which $q_{ab}^u = 0$. The $r^{uv}$ redundancy index is defined by the sum of the increase in the shortest paths in the network without both line sections *u* and *v* compared to the sum of the increase in the shortest paths in the network without line section *v*:

$$r^{uv} = q^{uv} - q^v = (c^{uv} - c) - (c^v - c) = c^{uv} - c^v \qquad (3)$$

By calculating $r^{uv}$ for all *v* line sections that are not identical with *u* and summing them up, one gets the total redundancy that line section *u* provides to line section *v*:

$$r^u = \sum_v r^{uv} = \sum_v (q^{uv} - q^v) = \sum_v (c^{uv} - c^v) \qquad (4)$$

This definition was introduced by Jenelius. [31]

Application on 1-Edge-Connected Graphs

It can be seen from the definition, that if such line section(s) are deleted from the graph that makes at least one station unreachable from the others, the value of both $q^u$ and $r^u$ becomes infinite. The railway network of Hungary has this property, which means that the graph describing it is a so-called 1-edge-connected graph. In several cases, by deleting only one line section from the undisrupted graph the graph will remain connected.

However, if two line sections are deleted, the number of reasonable results will rapidly decrease. If all these line sections were excluded from the calculations, only a few would remain and if only those line sections were excluded which give infinity as a result in that particular calculation, then different line sections would be taken into account for each $v$ line section, which would make the obtained $r^u$ values incomparable to each other.

Therefore, it is practical to use the reciprocals of the travel time and trip length values of the shortest paths. By changing the order in which the difference is calculated in the summation of (2), the redundancy index remains positive since longer distances mean shorter values in the reciprocal space.

By summing the values of the redundancy indices calculated in the reciprocal space for all $v$ line sections, one gets the total redundancy of a line section $u$:

$$\sum_v r_\ell^{uv'} = \sum_v (c_\ell^{v'} - c_\ell^{uv'}) = \sum_v \left( \sum_{\langle a,b \rangle} \frac{1}{\ell_{ab}^v} - \sum_{\langle a,b \rangle} \frac{1}{\ell_{ab}^{uv}} \right) \quad (5)$$

$$\sum_v r_t^{uv'} = \sum_v (c_t^{v'} - c_t^{uv'}) = \sum_v \left( \sum_{\langle a,b \rangle} \frac{1}{t_{ab}^v} - \sum_{\langle a,b \rangle} \frac{1}{t_{ab}^{uv}} \right) \quad (6)$$

However, it is more informative to normalise these values with values of the total trip length or the total travel time of the undisrupted network (which value is denoted by $c_\ell'$ and $c_t'$, respectively):

$$r_\ell^{u'} = \frac{\sum_v r_\ell^{uv'}}{c_\ell'} = \frac{\sum_v (c_\ell^{v'} - c_\ell^{uv'})}{c_\ell'} = \frac{\sum_v \left( \sum_{\langle a,b \rangle} \frac{1}{\ell_{ab}^v} - \sum_{\langle a,b \rangle} \frac{1}{\ell_{ab}^{uv}} \right)}{\sum_{\langle a,b \rangle} \frac{1}{\ell_{ab}^0}} \quad (7)$$

$$r_t^{u'} = \frac{\sum_v r_t^{uv'}}{c_t'} = \frac{\sum_v (c_t^{v'} - c_t^{uv'})}{c_t'} = \frac{\sum_v \left( \sum_{\langle a,b \rangle} \frac{1}{t_{ab}^v} - \sum_{\langle a,b \rangle} \frac{1}{t_{ab}^{uv}} \right)}{\sum_{\langle a,b \rangle} \frac{1}{t_{ab}^0}} \quad (8)$$

The $r^{u'}$ redundancy index is the total relative decrease in the reciprocal trip length or travel time for those shortest paths that do not pass through the line section $u$ in the undisrupted network but pass through it in the case of the disruption of line section $v$ with line section $u$ fixed for the calculation.

However, because of the definition, the redundancy values of the line sections calculated in a specific graph cannot be used to compare with values obtained for line sections in other graphs. They have meaning only in that specific graph that they were calculated for and also only relative to the redundancy values of other line sections.

## Results and Discussion

### Redistribution of the Network Flow

Paths with Minimal Trip Length

By calculating the shortest path for all ⟨a,b⟩ pairs of stations with minimal trip lengths in the networks with the four V0 alternatives, the ratio of paths passing through the line sections will differ from the values in the network without V0 (presented in subchapter *Flow in the network without the V0*) as some of the shortest paths will use the new railway routes.

The total network trip length is 10,047,606 km in the network without the V0 line, which value decreases if the V0 line (in either route) is introduced (see Table 1).

*Table 1.*
*The percentile decrease in the total network trip length and the percentile decrease in the number of paths between all pairs of stations with shortest trip lengths passing through the line section with the highest traffic.* [Made by the authors.]

|  | V0 route alternative | | | |
|---|---|---|---|---|
|  | #1 | #2 | #3 | #4 |
| The decrease in total network trip length (%) | 0.85 | 1.27 | 1.29 | 1.39 |
| The decrease in the percentile ratio of the number of paths passing through the line section with the heaviest traffic | 6.66 | 8.08 | 14.71 | 13.94 |

The alternative causing the most decrease and thus the best is alternative #4, which makes the total network trip length 139,573 km shorter. However, the alternative making the line section in the network without the V0 handling the heaviest traffic, the Ferencváros–Kelenföld line section containing the Southern Railway Bridge, to handle the less traffic possible is alternative #3. Alternative #4 is almost as good, both causing about a 14% decrease in the ratio of paths passing through the Southern Railway Bridge.

The change in the percentile ratio of paths passing through each line section can be seen in Figure 5 for the four alternative V0 routes.

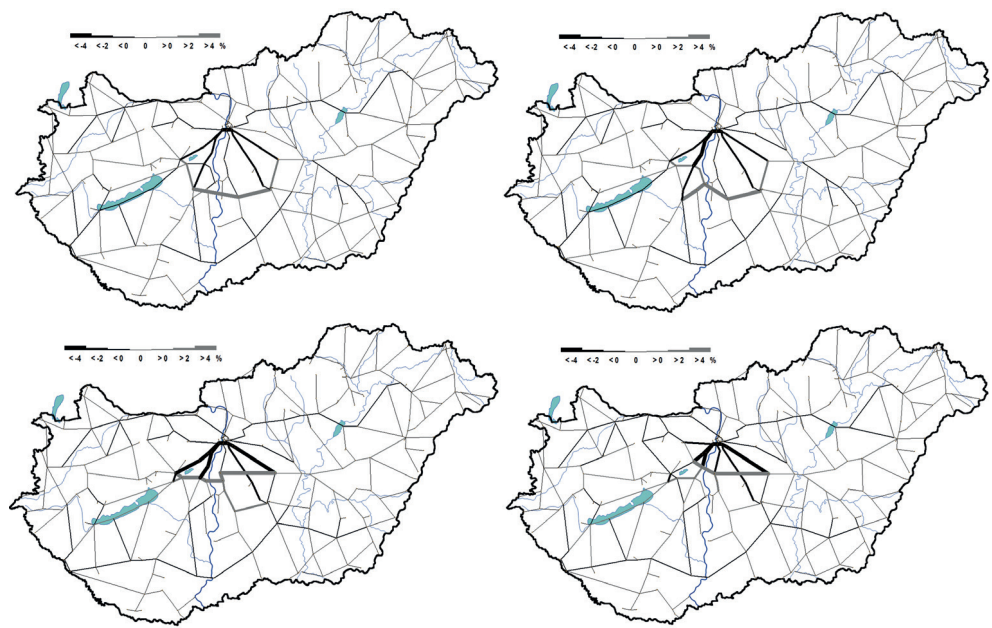

Figure 5. *The change in the percentile ratio of paths between all pairs of stations passing through each line section by introducing the four V0 alternatives (#1 top left, #2 top right, #3 bottom left, #4 bottom right) for minimal trip lengths.* [Made by the authors.]

In case of alternative #1, the ratio of the paths passing through the individual network elements decreases only for the Southern Railway Bridge with more than 4%: from 36.98% to 30.32%. The traffic of the radial main lines leading to Budapest (like lines 30a, 40a, 142, 100a) is affected less.

For alternative #2, the case is similar, only the traffic of line 40a between its crossing with V0 and Budapest decreases significantly.

Alternatives #3 and #4 are the best in reducing the traffic of the aforementioned main lines: these alternatives reroute so many paths that the traffic of existing lines decreases with more than 5%. If one of them has to be chosen, alternative #4 is the best choice because it not only reroutes the traffic effectively but also it is the shortest of the four.

Paths with Minimal Travel Time

The total network travel time is 6,834,569 minutes in the network without the V0 line, which value decreases if the V0 line (in either route) is introduced (see Table 2).

Table 2. *The percentile decrease in the total network travel time and the percentile decrease in the number of paths between all pairs of stations with the shortest travel times passing through the line section with the highest traffic.* [Made by the authors.]

|  | V0 route alternatives | | | |
|---|---|---|---|---|
|  | #1 | #2 | #3 | #4 |
| The decrease in total network travel time (%) | 1.23 | 1.03 | 0.82 | 1.56 |
| The decrease in the percentile ratio of the number of paths passing through the busiest line section | 2.49 | 2.41 | 2.73 | 14.29 |

The alternative causing the most decrease in the total network travel time and thus the best, similar to the case of trip lengths, is alternative #4: the total travel time decrease is 106,392.6 minutes. But in this case, the decrease this alternative causes in the traffic of the busiest line section is by far the highest among all four, more than 14%, which is six-time larger than for the other alternatives. This makes a clear indication that this is the best alternative not only for trip lengths but also for travel times.

The change in the percentile ratio of paths passing through each line section can be seen in Figure 6 for the four alternative V0 routes.

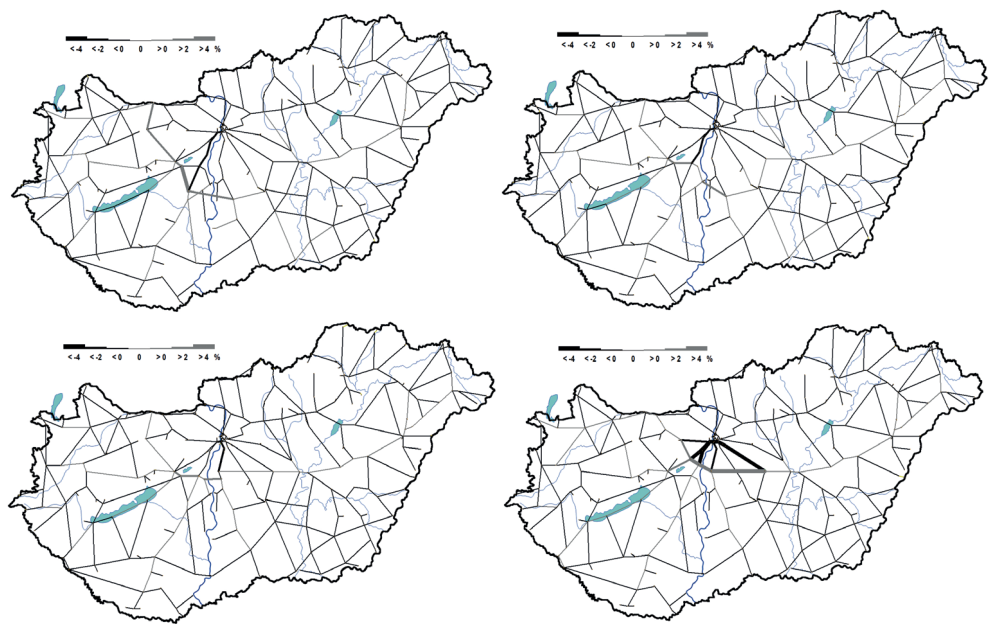

Figure 6. *The change in the percentile ratio of paths between all pairs of stations passing through each line section by introducing the four V0 alternatives (#1 top left, #2 top right, #3 bottom left, #4 bottom right) for minimal travel times.* [Made by the authors.]

As it can be seen, alternatives #1–#3 make only a small ratio of paths to reroute. The most affected line is 40a but the others are only affected by less than 2%. Alternative #4, however, makes the traffic of lines 1, 30a, 40a and 100a decrease dramatically and thus clearly reducing the traffic currently passing through Budapest. It reroutes the paths approaching Budapest, leading them around it on a faster way to their destination.

This means that alternative #4 is the best choice in simultaneously reducing the total network trip length, the total network travel time and rerouting as many paths as possible to bypass Budapest. A further advantage of this path is that it is sufficiently far from the capital of the country and the Southern Railway Bridge. [32]

## The Redundancy of V0

The redundancy value of the three-line sections with the existing Danube bridges and the (length and time-weighted) average redundancy of the V0 line alternatives were calculated for the network without the V0 line and for the four networks with the four V0 alternatives for both minimal trip length and minimal travel time paths. [33] The results can be seen in Figure 7.

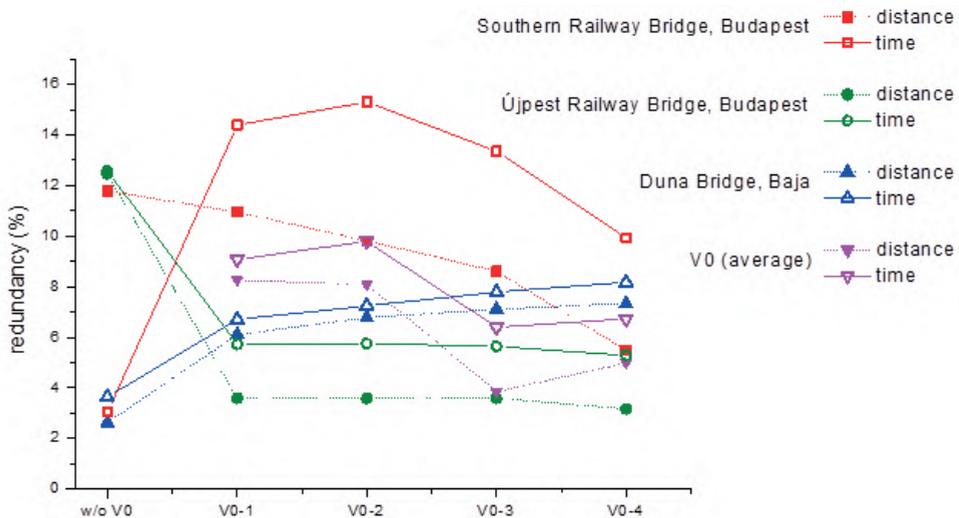

Figure 7. *The $r^u$' percentile redundancy values calculated using Eqs. (7) and (8), u being the line sections containing the Southern Railway Bridge (square), the Újpest Railway Bridge (circle), the Duna Bridge at Baja (upward triangle) and the Duna bridge of the V0 line (downward triangle). Each value is calculated for the network without the V0 line (w/o V0) and the four V0 alternatives (V0-1, V0-2, V0-3 and V0-4) for paths with the shortest trip length (closed symbol) and shortest travel time (open symbol), except for the V0 bridge for which the redundancy value cannot be calculated for the network without the V0 line.* [Made by the authors.]

Without V0

As it was shown by previous works, [24] [34] the Baja bridge has very little effect on the network as it is only used as a rerouting alternative for a few paths crossing the Southern bridge in the undisrupted network (see Table 3). No path passing through the Újpest bridge in the undisrupted network is rerouted through Baja. This is the case for paths with both minimal length and minimal time: the increase is so much that if this bridge is also disrupted, the further rerouting through the Újpest bridge makes a much smaller further increase in the length of the shortest paths and thus its redundancy is low.

The majority of paths passing through the Southern bridge in the undisrupted network are rerouted through the Újpest bridge on its disruption. The high redundancy value of the Újpest bridge is because the increase caused by this rerouting is relatively small related to the further increase caused by the rerouting through Baja, both for trip lengths and travel times.

The redundancy of the Southern Railway Bridge is very different for minimal trip lengths and travel times. For travel times, the increase caused by the rerouting through it from either of the other two bridges is so large, that the further increase caused by the rerouting when the Southern bridge is also disrupted, is relatively small. The case is very different for trip lengths. If a rerouted path has to pass through the Southern bridge instead of the Újpest bridge, it becomes only a few kilometres longer. Compared to this increase, rerouting through Baja on the disruption of the Southern bridge makes the paths much longer which increases its redundancy.

The rerouted paths passing through the Southern bridge in the undisrupted network are plotted in Figure 8.

Table 3. *The change of the number of paths passing through each Duna bridge on the disruption of one of the Duna bridges and the percentile change it means for that specific bridge compared to the number of paths passing thríough it in the undisrupted network.* [Made by the authors.]

| | Minimal distance | Change in the number of routes passing through the bridge | | |
|---|---|---|---|---|
| | | Southern | Újpest | Baja |
| Disruption of bridge | Southern | −15,603 (−100.0%) | +11,063 (+314.3%) | +3,907 (+330.17%) |
| | Újpest | +3,124 (+20.0%) | −3,520 (−100.0%) | 0 (0.0%) |
| | Baja | +1,015 (+6.5%) | 0 (0.0%) | −1,203 (−100.0%) |
| | Minimal time | Change in the number of routes passing through the bridge | | |
| | | Southern | Újpest | Baja |
| Disruption of bridge | Southern | −18,995 (−100.0%) | +13,105 (+1,378.0%) | +5,080 (+806.3%) |
| | Újpest | +457 (+2.4%) | −951 (−100.0%) | 0 (0.0%) |
| | Baja | +290 (+1.5%) | 0 (0.0%) | −630 (−100.0%) |

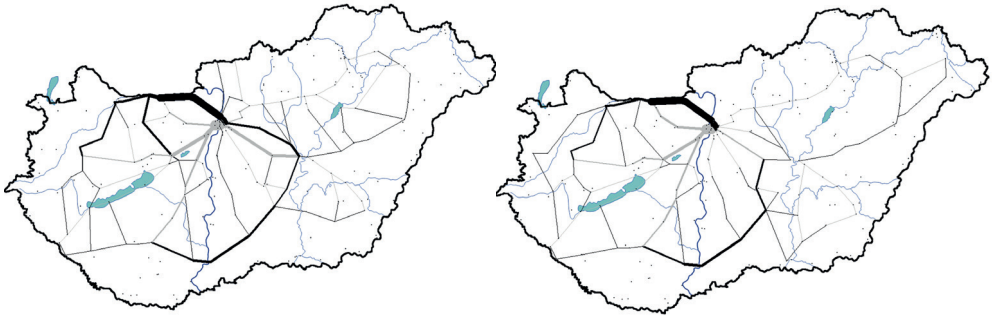

Figure 8. *The redistribution of paths with shortest trip length (left) and shortest travel time (right) on the disruption of the Southern Railway Bridge in the network without the V0 line. The thickness of the lines is proportional to the change in the number of paths passing through each line section: red standing for decrease and green standing for the increase* [Made by the authors.]

With V0

By introducing the V0 line, the redundancy values of the existing bridges converge. This is due to the role V0 ought to play in the railway network of Hungary: providing an alternative for the Southern Railway Bridge. As it is clear from the structure of the network, the disruption of the V0 bridge mostly makes the individual paths to reroute to their distribution as it was in the network without the V0 line (see Figure 9). Similarly, on the disruption of the Southern bridge, the vast majority of the lines is rerouted through the V0 bridge (and also the Újpest bridge). As it is clear from Figure 7, alternative #4 makes the network to be the most balanced in the view of redundancy as the difference between the values of the Southern bridge and the V0 bridge is the smallest for this alternative.

According to the traffic flow values (Table 4), the ratios of the rerouted paths between the three older bridges remain approximately the same, but the newly introduced V0 bridge becomes the main rerouting alternative for all other bridges, even some paths passing through the Újpest bridge in the undisrupted network. Paths passing through the Újpest bridge and the Baja bridge in the undisrupted network are never rerouted on the other one, as was the case in the network without the V0 line.

The change in the flow values in Table 4 clearly shows that V0 becomes the most important rerouting alternative. This is because the disruption of the V0 bridge results in significantly longer paths which, combined with the high number of paths passing through it leads to a high redundancy value. As the spatial distribution of the Danube bridges become more even on the introduction of the V0 line, one alternative route is always significantly better than the others for a chosen path. This also means that there is always a better rerouting path in the network with the V0 line than in the present network without V0 and thus the substitution of the Danube bridges can be handled much more sufficiently making the railway network of the country less vulnerable. [35]

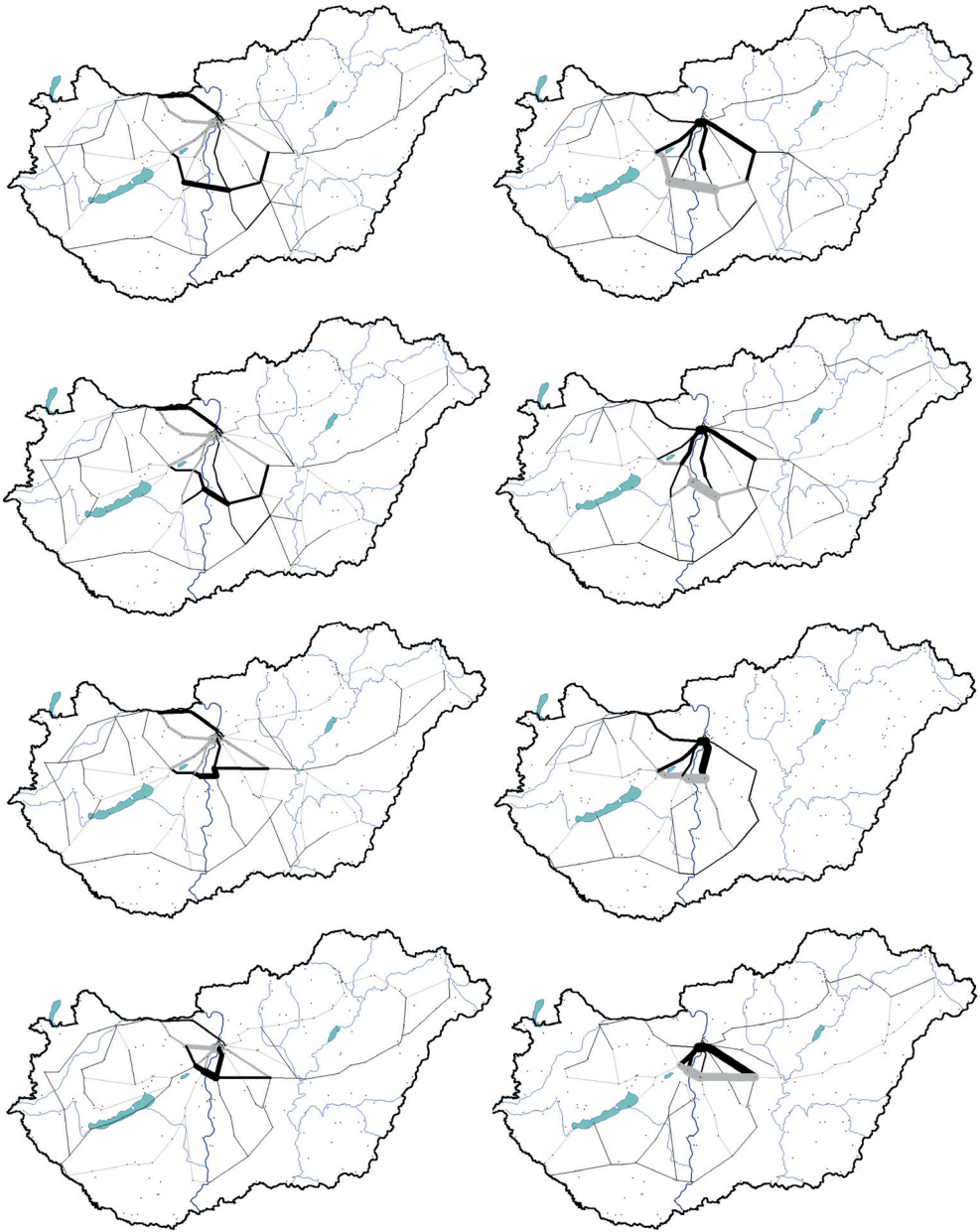

Figure 9. *The redistribution of paths with shortest travel time on the disruption of the Southern Railway Bridge (left) and the Duna bridge on the V0 line (right) in the network with the V0 line (from top to bottom V0 alternative #1, #2, #3 and #4). The thickness of the lines is proportional to the change in the number of paths passing through each line section: red standing for decrease and green standing for the increase.*
[Made by the authors.]

Table 4. *The change of the number of paths passing through each Duna bridge on the disruption of one of the Duna bridges and the percentile change it means for that specific bridge compared to the number of paths passing through it in the undisrupted network.* [Made by the authors.]

| Alternative #1 min. distance | | Change in the number of routes passing through the bridge | | | |
|---|---|---|---|---|---|
| | | Southern | Újpest | Baja | V0 |
| Disruption of bridge | Southern | −12,794 (−100.0%) | +6527 (+185.4%) | −32 (−3.8%) | +5731 (+181.1%) |
| | Újpest | +3,124 (+24.4%) | −3520 (−100.0%) | 0 (0.0%) | 0 (0.0%) |
| | Baja | −27 (−0.2%) | 0 (0.0%) | −847 (−100.0%) | +686 (+21.7%) |
| | V0 | +2809 (+22.0%) | 0 (0.0%) | +356 (+42.0%) | −3165 (−100.0%) |

| Alternative #1 min. time | | Change in the number of routes passing through the bridge | | | |
|---|---|---|---|---|---|
| | | Southern | Újpest | Baja | V0 |
| Disruption of bridge | Southern | −17,835 (−100.0%) | 6939 (+729.7%) | +240 (+50.0%) | +9838 (+751.0%) |
| | Újpest | +429 (+2.4%) | −951 (−100.0%) | 0 (0.0%) | +28 (+2.1%) |
| | Baja | −75 (−0.4%) | 0 (0.0%) | −480 (−100.0%) | +215 (+16.4%) |
| | V0 | +1160 (+6.5%) | 0 (0.0%) | +150 (+31.3%) | −1310 (−100.0%) |

| Alternative #2 min. distance | | Change in the number of routes passing through the bridge | | | |
|---|---|---|---|---|---|
| | | Southern | Újpest | Baja | V0 |
| Disruption of bridge | Southern | −12,195 (−100.0%) | +5,441 (+154.6%) | −50 (−10.9%) | +6,276 (+151.2%) |
| | Újpest | 3,124 (+25.6%) | −3,520 (−100.0%) | 0 (0.0%) | 0 (0.0%) |
| | Baja | −72 (−0.6%) | 0 (0.0%) | −460 (−100.0%) | +344 (+8.3%) |
| | V0 | +3,408 (+27.9%) | 0 (0.0%) | +743 (+161.5%) | −4,151 (−100.0%) |

| Alternative #2 min. time | | Change in the number of routes passing through the bridge | | | |
|---|---|---|---|---|---|
| | | Southern | Újpest | Baja | V0 |
| Disruption of bridge | Southern | −17,892 (−100.0%) | +7,288 (+766.4%) | +244 (+50.8%) | +9,710 (+774.9%) |
| | Újpest | +441 (+2.5%) | −951 (−100.0%) | 0 (0.0%) | +16 (+1.3%) |
| | Baja | −79 (−0.4%) | 0 (0.0%) | −480 (−100.0%) | +219 (+17.5%) |
| | V0 | +1,109 (+6.2%) | 0 (0.0%) | +144 (+30.0%) | −1,253 (−100.0%) |

| Alternative #3 min. distance | | Change in the number of routes passing through the bridge | | | |
|---|---|---|---|---|---|
| | | Southern | Újpest | Baja | V0 |
| Disruption of bridge | Southern | −9,398 (−100.0%) | +3,702 (+105.5%) | −18 (−1.8%) | +5,516 (+85.8%) |
| | Újpest | +3,134 (+33.3%) | −3,508 (−100.0%) | 0 (0.0%) | −22 (−0.3%) |
| | Baja | −31 (−0.3%) | 0 (0.0%) | −1,005 (−100.0%) | +832 (+12.9%) |
| | V0 | +6,205 (+66.0%) | +12 (+0.3%) | +198 (+19.7%) | −6,431 (−100.0%) |

| Alternative #3 min. time | | Change in the number of routes passing through the bridge | | | |
|---|---|---|---|---|---|
| | | Southern | Újpest | Baja | V0 |
| Disruption of bridge | Southern | −18,128 (−100.0%) | +6,696 (+704.1%) | +2 (+0.4%) | +10,918 (+1134.9%) |
| | Újpest | +457 (+2.5%) | −951 (−100.0%) | 0 (0.0%) | 0 (0.0%) |
| | Baja | −24 (−0.1%) | 0 (0.0%) | −535 (−100.0%) | +219 (+22.8%) |
| | V0 | +868 (+17.6%) | 0 (0.0%) | +94 (+17.6%) | −962 (−100.0%) |

| Alternative #4 min. distance | | Change in the number of routes passing through the bridge | | | |
|---|---|---|---|---|---|
| | | Southern | Újpest | Baja | V0 |
| Disruption of bridge | Southern | −9,723 (−100.0%) | +3,089 (+89.1%) | −18 (−1.7%) | +6,464 (+105.7%) |
| | Újpest | +3,022 (+31.1%) | −3,466 (−100.0%) | 0 (0.0%) | +48 (+0.8%) |
| | Baja | −84 (−0.9%) | 0 (0.0%) | −1,039 (−100.0%) | +919 (+15.0%) |
| | V0 | +5,880 (+60.5%) | +54 (+1.6%) | +164 (+15.8%) | −6,114 (−100.0%) |

| Alternative #4 min. time | | Change in the number of routes passing through the bridge | | | |
|---|---|---|---|---|---|
| | | Southern | Újpest | Baja | V0 |
| Disruption of bridge | Southern | −12,858 (−100.0%) | +2,442 (+256.8%) | −155 (−28.9%) | +10,125 (+162.3%) |
| | Újpest | +409 (+3.2%) | −951 (−100.0%) | 0 (0.0%) | +48 (+0.8%) |
| | Baja | −164 (−1.3%) | 0 (0.0%) | −537 (−100.0%) | +353 (+5.7%) |
| | V0 | +6,137 (+47.7%) | 0 (0.0%) | +93 (+17.3%) | −6,238 (−100.0%) |

## Conclusions

We modelled the effect of the long-planned V0 railway line on the railway network of Hungary using a weighted directed graph. Four of the proposed alternative routes were analysed. It is clear from our results that alternative #4, the one not too close to and not too far from Budapest leads to the best results. As it is not too far from the Southern Railway Bridge, it can provide a reasonable alternative route to it, and as it is not too close to it, the rerouting is not limited exclusively to the paths passing through the Southern bridge.

Regarding travel times, alternatives #3 and #4 provide the best redistribution of paths for paths with the shortest possible length, but for paths with minimal travel times, alternative #4 is the only case in which the Southern bridge is effectively disencumbered.

The V0 line provides high redundancy for the network as many routes are rerouted through it on the disruption of the other Danube bridges and its disruption leads to a significant further increase in the length of these paths. V0 also increases the redundancy of the other three bridges: as numerous paths pass through it in the undisrupted network (particularly in the case of alternative #4), on the rerouting of these paths, they have to pass through one of the other bridges, and their simultaneous disruption increases the trip length or the travel time further with a significant amount.

This means that the V0 line is not only necessary for the railway network of Hungary to make the existing bridges (especially the Southern Railway Bridge in Budapest) handle moderate traffic but also for the redundancy it provides in the case of the disruption of any other Danube bridge.